\documentclass[aps,prb,twocolumn,superscriptaddress]{revtex4-2}

\usepackage{amsmath}
\usepackage{mathtools} 
\usepackage{amssymb}
\usepackage{amsthm}
\usepackage{bm,upgreek} 
\usepackage{upgreek} 
\usepackage{xcolor}
\usepackage{graphicx}
\usepackage{epstopdf}
\usepackage[colorlinks,linkcolor=blue,anchorcolor=blue,citecolor=blue,urlcolor=blue]{hyperref}

\def\eqa{\begin{eqnarray}}
\def\eea{\end{eqnarray}}
\newcommand{\eq}{\begin{equation}}
\newcommand{\ee}{\end{equation}}
\newcommand{\nn}{\nonumber\\}

\newcommand{\<}{\langle}
\renewcommand{\>}{\rangle}

\newcommand{\del}{\delta}
\newcommand{\Del}{\Delta}
\newcommand{\eps}{\epsilon}
\newcommand{\veps}{\varepsilon}

\newcommand{\cD}{ {\cal D} }
\newcommand{\cE}{ {\cal E} }

\newcommand{\cN}{ {\cal N} }
\newcommand{\cP}{ {\cal P} }

\usepackage{color}

\begin{document}

\title{Transition from band insulator to Mott insulator and formation of local moment in  half-filled ionic SU($N$) Hubbard model}

\author{Shan-Yue Wang}
\affiliation{National Laboratory of Solid State Microstructures and School of Physics, Nanjing
	University, Nanjing, 210093, China}

\author{Da Wang}
\affiliation{National Laboratory of Solid State Microstructures and School of Physics, Nanjing
	University, Nanjing, 210093, China}
\affiliation{Collaborative Innovation Center of Advanced Microstructures, Nanjing University, Nanjing 210093, China}

\author{Qiang-Hua Wang}
\affiliation{National Laboratory of Solid State Microstructures and School of Physics, Nanjing
University, Nanjing, 210093, China}
\affiliation{Collaborative Innovation Center of Advanced Microstructures, Nanjing University, Nanjing 210093, China}

\begin{abstract}
We investigate {the local moment formation in}
the half-filled SU($N$) Hubbard model under a staggered ionic potential.
As the Hubbard $U$ increases, the charge fluctuations are suppressed and eventually frozen when $U$ is above a critical value $U_c$, marking the development of well-defined local moment with integer $m$ fermions on the A-sublattice and $(N-m)$ fermions on the B-sublattice, respectively. We obtain an analytical solution for $U_c$ {for the paramagnetic ground state} within the variational Gutzwiller approximation and renormalized mean field theory.
For large $N$, $U_c$ is found to depend on $N$ linearly with fixed $m/N$, but sublinearly with fixed $m$.
The local moment formation is accompanied by a peculiar phase transition from the band insulator to the Mott insulator, where the ionic potential and quasiparticle weight are renormalized to zero simultaneously.
Inside the Mott phase, the low energy physics is described by the SU($N$) Heisenberg model with conjugate representations, which is widely studied in the literature.
\end{abstract}


\maketitle

\section{Introduction}

Quantum spin models are a class of physical models describing ``spins'' or ``local moments'' which originate from strong correlations between fermions (e.g. electrons or cold atoms) such that the ``charge'' (fermion number) degrees of freedom are frozen \cite{Anderson_PR_1961}.
For instance, the Heisenberg model is a low energy description of the Hubbard model only when the Hubbard $U$ is large enough to drive the system into the Mott insulating phase \cite{Brinkman_PRB_1970, Imada_RMP_1998}.
In the literature, the SU(2) Heisenberg model has been generalized to the SU($N$) case \cite{Affleck_PRL_1985} with the spin operators satisfying the SU($N$) algebra.
The SU($N$) Heisenberg models provide a vast area to explore many new phenomena \cite{Affleck_PRB_1988, Marston_PRB_1989, Arovas_PRB_1988, Read_PRL_1989, Read_NPB_1989, Read_PRB_1990, Harada_PRL_2003, Assaad_PRB_2005, Kawashima_PRL_2007, Arovas_PRB_2008, Xu_PRB_2008, Wu_P_2010, Beach_PRB_2009, Lou_PRB_2009, Rachel_PRB_2009, Kaul_PRL_2012, Corboz_PRB_2012, Harada_PRB_2013, Nataf_PRL_2014, Dufour_PRB_2015, Okubo_PRB_2015, Suzuki_PRB_2015, Nataf_PRB_2016, Demidio_PRB_2016, Nataf_PRB_2018, Kim_PRB_2019}.
Among different SU($N$) representations for the local spins, a conjugate representation with $m$ fermions on A-sublattice and $(N-m)$ fermions on B-sublattice \cite{Affleck_PRL_1985} is mostly studied.
It may support a generalized Neel order: for instance, the first $m$ and remaining $(N-m)$ flavors are occupied on the two sublattices, respectively.
Hence, the SU($N$) symmetry is broken into $\mathrm{SU}(m) \times \mathrm{SU}(N-m)$.
The gapless fluctuations above this Neel order fall into the Grassmannian manifold $\mathrm{Gr}(N,m) = \mathrm{U} (N)/[\mathrm{U} (m)\times \mathrm{U}(N-m)]$ \cite{MacFarlane_PLB_1979, Hikami_PTP_1980, Duerksen_PRD_1981, Maharana_AIPT_1983}, which is reduced to the $N$-component Ginzburg-Landau theory \cite{Halperin_PRL_1974} or equivalently the famous CP$^{N-1}$ model in the special case of $m=1$.

One early motivation for doing the SU($N$) generalization of Heisenberg model is to perform  $1/N$ expansion around the saddle point at $N=\infty$, providing a controllable way to reach the SU(2) model \cite{Auerbach_book_1994}.
However, physically speaking, the SU($N$) spin models should derive from the SU($N$) Hubbard model in the limit that charge fluctuations are completely frozen. The SU($N$) Hubbard model is widely studied,\cite{Lu_PRB_1994, Wu_PRL_2003, Honerkamp_PRL_2004, Buchta_PRB_2007, Hung_PRB_2011, Cai_PRL_2013, Cai_PRB_2013, Lang_PRL_2013, Wang_PRL_2014, Zhou_PRB_2014, Zhou_PRB_2016, Zhou_PRB_2017, Zhou_PRB_2018, Wang_PRB_2019_QMC}, and is now within experimental reach, thanks to the fast technical development, mostly in the field of cold atoms \cite{Bloch_RMP_2008,Bloch_NP_2012, DeSalvo_PRL_2010, Taie_PRL_2010, Krauser_NP_2012, Taie_NP_2012, Zhang_S_2014, Cazalilla_RPP_2014, Hart_N_2015, Laflamme_AP_2016, Hofrichter_PRX_2016, Ozawa_PRL_2018, Sonderhouse_NP_2020, Taie_A_2020}.
This brings the large-$N$ model to life, but not just a gedanken model, and opens up a new field in the study of the finite but large $N$ versions of such models, in the search for novel quantum spin states.

However, it is important to ask under what condition is the system aptly described by the quantum spin model for which the local moments have to be well established.
In our previous work, we have proposed to add a staggered ionic potential to the SU($N$) Hubbard model \cite{Wang_PRB_2019}.
In this work, we will examine the specific conditions for the Hubbard $U$ and ionic potential $V$ under which the local moments can exist, with immediate relevance to experimental realization.
We develop and apply an SU($N$)-symmetric renormalized mean field theory (RMFT) based on the variational approach of the Gutzwiller projection approximation \cite{Gutzwiller_PRL_1963, Gutzwiller_PR_1965, Brinkman_PRB_1970, Vollhardt_RMP_1984, Bunemann_PRB_1998, Wang_PRB_2006, Edegger_AP_2007}.
The RMFT developed here may be applied or extended straightforwardly for general models with a large number of fermion flavors subject to any internal symmetry.
For the ionic SU($N$) Hubbard model, we find the local moments, with quantized integer $m$ fermions on the A-sublattice and $(N-m)$ fermions on the B-sublattice, are well established when $U$ is above a critical value $U_c$, which depends on $N$, $m$, and the ionic potential $V$.
For large $N$, $U_c$ is found to depend on $N$ linearly with fixed $m/N$, but sublinearly with fixed $m$.
In addition, the local moment formation is accompanied by a peculiar transition from the band insulator to the Mott insulator \cite{Garg_PRL_2006, Kancharla_PRL_2007}, at which the ionic potential and quasiparticle weight are renormalized to zero simultaneously.
Finally, we show that the low energy physics of local moments is described by the widely studied SU($N$) quantum spin model inside the Mott insulating phase. Our results shed light on the realization of such models in, e.g., cold atoms.

\section{SU($N$) Hubbard model in an ionic potential}

Th ionic SU($N$) Hubbard model we consider is described by the Hamiltonian $H=H_t+H_U$, with
\eqa
H_t&=&-t\sum_{\<ij\>,a} [c_{a}^\dag(i) c_{a}(j)+{\rm H.c.}] ,
\eea
and
\eqa
H_U&=&\frac{U}{2}\sum_i\left[\hat{n}(i)-\frac{N}{2}\right]^2 + V\sum_i (-1)^i \hat{n}(i) ,
\eea
where $i$ labels the lattice site, $\<ij\>$ denotes a nearest-neighbor bond, $a=1, 2, \cdots, N$ labels the flavor of the fermions, $\hat{n}(i)=\sum_a \hat{n}_a(i)=\sum_a c_{a}^\dag(i) c_{a}(i)$ is the local density operator, $U$ denotes the Hubbard interaction, and $V$ is the staggered ionic potential.
The model is clearly SU($N$)-symmetric in the internal flavor space.
In real space, it breaks the translational symmetry, since A- and B-sublattices are distinct.
However, a particle-hole transformation $c_{ia}\to(-1)^ic_{ia}^\dag$ interchanges these sublattices, so the system is invariant under A-B sublattice transformation combined with the particle-hole transformation.
{As a result, the charge density is exactly staggered and the system is at half filling on average. Such a symmetry can be used to reduce the variational parameters, as can be seen in later discussions.}

The $H_U$-term can be rewritten as
\eqa H_U= \frac{U}{2} \sum_i \left[\hat{n}(i)-n_{0}(i)\right]^2 , \eea
where $n_{0}(i)=\frac{N}{2}-(-1)^i\frac{V}{U}$ may be understood as the local ground charge tunable continuously by the staggered gate voltage $V$.
We ask whether $\hat{n}(i)$ can be quantized to integers $m$ ($0<m<N$) on the A-sublattice and $(N-m)$ on the B-sublattice, i.e., forming local moments, called $m$-tuple moments, for large enough $U$.
The concept of local moment is a natural generalization of the SU(2) case for which only one kind of local moment with $m=1$ is possible.
Here, however, the states with different $m$-tuple moments should belong to different Mott insulating states.
On the other hand, in the limit of $U=0$, a nonzero $V$ always yields a band insulator.
For large enough $U$, the system is expected to enter different Mott insulating states labeled by $m$.
Whether these $m$-tuple moments exist and how to describe these band-to-Mott insulator transitions are the main concerns of the present work.
{To answer these questions clearly, and for simplicity, we shall focus on the paramagnetic case in the following.}

\section{SU($N$)-symmetric Gutzwiller projection approximation and RMFT}

Local moment formation is beyond any Hartree-Fock mean field description.
We employ the standard Gutzwiller projection approximation to treat the correlation effect.
{ The SU(2) version of such a theory has been applied widely
\cite{Gutzwiller_PR_1965, Brinkman_PRB_1970, Vollhardt_RMP_1984, Wang_PRB_2006, Edegger_AP_2007},
and will be extended here} to the SU($N$)-symmetric case in which all the $N$-flavors are equivalent. For sufficient generality, we present the theory for an arbitrary case of the applied potential in this section, and will specify the ionic potential in the next section.

Specifically, we consider a variational theory with the following trial wave function,
\eqa |\psi\> = \hat{\cP} |\psi_0\>, \eea
where $|\psi_0\>$ is the ground state of a free variational Hamiltonian to be specified, and $\hat{\cP}$ is the Gutzwiller projection operator in the grand canonical ensemble
\eqa \hat{\cP} = \Pi_i \hat{\cP}_i, \ \ \hat{\cP}_i = \sum_{k=0}^{N} \eta_k(i) y_i^k \hat{Q}_k(i) , \label{eq:proj}\eea
where $\hat{Q}_k(i)$ is the projection operator for the $k$-tuple state (with $k$-fermions),
\eqa \hat{Q}_k(i) = \sum_{S=\{a_\ell|\ell=1,\cdots,k\}} \prod_{b \in S} \hat{n}_{b}(i) \prod_{b \notin S} [1-\hat{n}_{b}(i)].\eea
Clearly, in the absence of projection, we have $\hat{\cP}_i=\sum_k \hat{Q}_k(i)=1$.
The idea of the Gutzwiller projection is to reassign weights to the basis states, and this is how correlations (at least the local ones) can be captured.
Here, the weight for the $k$-tuple is assumed to be
\eqa
\eta_k(i) y_i^k =\exp(-g_ik^2 +k\ln y_i)
\eea
where $g_i$ is the site-dependent Gutzwiller projection parameter to punish multi-fermion occupations (but can be chosen to be uniform in our case).
In the spirit of density functional theory \cite{Hohenberg_PR_1964}, the ground state energy is a unique functional of the density distribution.
Therefore we have introduced a fugacity $y_i$ to maintain the fermion density before and after projection in the grand canonical ensemble \cite{Wang_PRB_2006}. The fermion density on each site before projection is
\eqa Nf_i = \<\hat{n}(i)\>_0 = \sum_k k q_{k0}(i) , \label{eq:f-before}\eea
where $f_i$ is the average occupation number per flavor, and $\<\cdot\>_0$ indicates the average performed with respect to $|\psi_0\>$. We have defined $q_{k0}(i)$ as the average of $\hat{Q}_k(i)$ in the unprojected state,
\eqa q_{k0}(i)=\<\hat{Q}_k(i)\>_0 =C_N^k f^k(i) [1-f(i)]^{N-k} , \label{eq:qk0}\eea
where $C_N^k$ is the combinatorial factor. After projection, we still require
\eqa Nf_i = \<\hat{n}(i)\> = \sum_k k q_{k}(i) , \label{eq:f-after}\eea
with
\eqa q_{k}(i)=\<\hat{Q}_k(i)\> = \frac{\<\hat{\cP} \hat{Q}_k(i) \hat{\cP}\>_0}{\<\hat{\cP} \hat{\cP}\>_0} ,\eea
where $\<\cdot\>$ denotes average with respect to $|\psi\>$.
Exact evaluation on the right hand side is difficult.
To make analytical progress, we resort to the usual Gutzwiller approximation \cite{Gutzwiller_PR_1965}: the projection operator unrelated to the target operator under average can be Wick-contracted separately.
This approximation can be shown to be exact in infinite dimensions { \cite{Gebhard1990} for general on-site interactions \cite{Bunemann_PRB_1998} }, and turns out to work satisfactorily in finite dimensions \cite{Vollhardt_RMP_1984,Bunemann_PRB_1998,Edegger_AP_2007}.
Under the Gutzwiller approximation, we have
\eqa q_k(i) = \frac{\<\hat{\cP}_i \hat{Q}_k(i)\hat{\cP}_i\>_0}{\<\hat{\cP}_i \hat{\cP}_i\>_0}.\eea
Note the projection operator $\cP$ is simplified to $\cP_i$.
After substituting $\hat{\cP}_i$ in Eq.~\ref{eq:proj}, we obtain
\eqa q_k(i) = \frac{\eta_k^2(i) y_i^{2k} q_{k0}(i)}{\cD_i}, \,\, \cD_i=\sum_k \eta_k^2(i) y_i^{2k} q_{k0}(i) ,\label{eq:qk}\eea
where we have used $\hat{Q}_k^2(i)=\hat{Q}_k(i)$ as a property of the projection operator $\hat{Q}_k(i)$.
The fugacity $y_i$ (or $\ln y_i$ in practice) is then tuned to satisfy the density restriction Eq.~\ref{eq:f-after}.

After obtaining all of $q_{k0}(i)$ and $q_k(i)$, we are in a position to evaluate the variational energy in the projected state under the Gutzwiller approximation.
The local charging energy is obtained most straightforwardly, given the fact that the total charge operator and the projectors $\hat{Q}_k(i)$ share the same local basis states as eigenstates,
\eqa E_U=\< H_U \> = \frac{U}{2}\sum_{i}\sum_{k=0}^N \left[k-n_0(i)\right]^2 q_k(i).\eea
The kinetic energy is slightly more difficult to evaluate.
Since the fermion hopping involves two sites, we need to keep two projectors, say,  $\hat{\cP}_i $ and $\hat{\cP}_j$ in the hopping on the $\<ij\>$ bond,
\eqa  \<c_a^\dag(i) c_a(j)\> = \frac{\<\hat{\cP}_i c_a^\dagger(i) \hat{\cP}_i \hat{\cP}_j c_a(j) \hat{\cP}_j\>_0}{\<\hat{\cP}_i^2\hat{\cP}_j^2\>_0}. \label{eq:cij}\eea
Since the fermion operator is self-projective, we need to remove over projections before taking the quantum average.
For a given site $i$, we observe that
\eqa \hat{\cP}_i c_a(i) \hat{\cP}_i &=& \sum_k [\eta_k(i) y_i^k \hat{Q}_k(i)] c_a(i) [\eta_{k+1}(i) y_i^{k+1} \hat{Q}_{k+1}(i)] \nn &\equiv& \sum_k [\eta_k(i) \eta_{k+1}(i) y_i^{2k+1} \hat{Q}_k^{\hat{a}}(i)] c_a(i),\eea
where we have defined a partial projection operator $Q_{k}^{\hat{a}}(i)$ for $k$ fermions in the local Fock space excluding flavor $a$,
\begin{align} \hat{Q}_k^{\hat{a}}(i) = \sum_{
S=\{a_\ell|\ell=1,\cdots,k; a_\ell\ne a\}} \prod_{b\in S}\hat{n}_b(i) \prod_{b\notin S,b\neq a} [1-\hat{n}_b(i)]. \end{align}
Its average in $|\psi_0\>$ is evaluated to be
\eqa q_{k0}^{\hat{a}}(i)=\<\hat{Q}_{k}^{\hat{a}}(i)\>_0 = C_{N-1}^{k}f_i^{k}(1-f_i)^{N-1-k} ,\label{eq:qk0a}\eea
for any flavor $a$ in our SU($N$)-symmetric case.
Inserting the above relations in Eq.~\ref{eq:cij}, we obtain
\eqa \<c_a^\dag(i) c_a(j)\> = g_{t}(i,j) \<c_a^\dag(i) c_a(j)\>_0, \eea
where $g_{t}(i,j)= z(i)z(j)$ is the renormalization of the hopping by the projection, and
\eqa z(i) &=& \frac{\sum_k \eta_k(i)\eta_{k+1}(i)y_i^{2k+1}q_{k0}^{\hat{a}}(i)}{\cD_i} \nn
&=&\sum_{k=0}^{N-1} q_{k0}^{\hat{a}}(i) \sqrt{\frac{q_k(i) q_{k+1}(i)}{q_{k0}(i)q_{k+1,0}(i)}}.\eea
In the second line we have used Eq.(\ref{eq:qk}) to trade $\eta_k(i)y_i^k/\sqrt{\cD_i}$ for $\sqrt{q_k(i)/q_{k0}(i)}$.

Combining the potential and kinetic energies, we obtain the total variational energy $E$ in the projected state,
\begin{align} E = -t\sum_{\<ij\>a}g_{t}(i,j)\chi_{ij0} + \frac{U}{2}\sum_i \sum_{k=0}^N \left[k-n_0(i)\right]^2 q_k(i), \label{eq:E} \end{align}
where $\chi_{ij0}= \left\< c_a^\dag(i) c_a(j)+{\rm H.c.}\right\>_0$ is the average of hopping operator in the unprojected state.
This energy is understood as a functional of (i) the fermion density $f_i$ which in turn depends on the trial wavefunction $|\psi_0\>$, and (ii) the Gutzwiller projection parameter $g_i$.
The fugacity parameters $y_i$ are taken as Lagrange multipliers that are eliminated by forcing the invariance of the local fermion density under the projection.
The variational Gutzwiller approximation is closely related to the RMFT.
Minimizing $E$ with respect to $|\psi_0\>$, i.e., $\del E/\del \<\psi_0| = 0$, with fixed fermion density $Nf_i$, we obtain
\eqa H_{\rm RMFT}|\psi_0\> = \cE|\psi_0\>,\eea
where $\cE$ is introduced as the Lagrange multiplier enforcing normalization of the wave function, and $H_{\rm RMFT}$ is a free Hamiltonian yet encoded with the renormalization effect from the Gutzwiller projection,
\begin{align} H_{\rm RMFT} =-t\sum_{\<ij\>a}g_{t}(i,j) [c_a^\dag(i) c_a(j) + {\rm H.c.}]
  -\sum_{i}\mu_i \hat{n}(i),
\end{align}
where the variational local chemical potential $\mu_i$ is introduced to enforce $Nf_i=\<\hat{n}(i)\>_0$.
It can be shown that the single-particle spectrum of $H_{\rm RMFT}$ is just the  quasipaticle excitation spectrum beyond the correlated variational ground state, with the quasiparticle weight renormalized by $g_t$ \cite{Wang_PRB_2006}.


\section{Application to the ionic SU($N$) Hubbard model}

In this section, we apply the variational Gutzwiller approximation and RMFT developed in the previous section to the ionic SU($N$) Hubbard model in our interest.

\subsection{General formalism}
Due to the particle-hole and sublattice symmetries, and without involving further symmetry breaking, we only have to specify the fermion density (per flavor) $f$ and the Gutzwiller parameter $g$ on the A-sublattice.
Correspondingly, we can replace $f\to 1-f$, $k\to N-k$, $n_0\to N-n_0$, {etc}., to obtain the relevant quantities on the B-sublattice, while $g$ remains the same.
Under these simplifications, $\chi_{ij0}$ and $g_t(i,j)$ become bond-independent, denoted as $\chi_0$ and $g_t$, respectively. In particular, $g_t$ is now given by
\eqa g_t = \left(\sum_k q_{k0}^{\hat{a}}\sqrt{\frac{q_k q_{k+1}}{q_{k0}q_{k+1,0}}}\right)^2. \label{eq:gt}\eea
Due to the presence of an ionic potential, we choose $\mu_i=-(-1)^ig_t\Delta_c$ in $H_{\rm RMFT}$ to write
\eqa H_{\rm RMFT} &=& -g_t t\sum_{\<ij\>a} [c_a^\dag(i) c_a(j) + {\rm H.c.}] \nonumber\\ &&+g_t\Delta_c\sum_{i}(-)^i \hat{n}(i). \label{eq:Hrmft} \eea
Under such a parametrization, $g_t$ is a global factor renormalizing the effective bandwidth and quasiparticle excitations.
The unprojected ground state $|\psi_0\>$, subsequently the fermion density $f$, and the average hopping $\chi_0$, only depend on $\Del_c$. From $H_{\rm RMFT}$ and after some algebra, we obtain
\begin{align}
&f = \frac12 \int d\veps \rho(\veps) \left(1 - \frac{\Del_c}{\sqrt{\veps^2+\Del_c^2}}\right). \\
&\zeta t\chi_0 = \int d\veps \rho(\veps) \frac{\veps^2}{\sqrt{\veps^2+\Del_c^2}}.
\end{align}
Here $\zeta$ is the coordination number, and $\rho(\veps)$ is the density of states. As an illustrative example, we consider the Bethe lattice, for which $\rho(\veps) = (4/\pi W)\sqrt{ 1-{4\veps^2}/{W^2} }$, with $W$ the bandwidth, giving rise to
\begin{align}
&f = \frac12-\frac12\frac{\tilde{\Delta}}{\sqrt{\tilde{\Delta}^2+1}} \,_2F_1\left( \frac12, \frac32; 2; \frac{1}{\tilde{\Delta}^2+1} \right) \label{eq:Bethe-f},\\
&\zeta t\chi_0=\frac{W}{4} \,_2F_1\left( \frac12, \frac32; 3; \frac{1}{\tilde{\Delta}^2+1} \right), \label{eq:Bethe-chi0}
\end{align}
where $\tilde{\Delta}=2\Delta_c/W$ and $_2F_1$ is the standard hypergeometric function.

In practice, for a given $f$, we construct $q_{k0}$ (Eq.~\ref{eq:qk0}), $q_{k0}^{\hat{a}}$ (Eq.~\ref{eq:qk0a}), and $q_k$ (Eq.~\ref{eq:f-after} by tuning $y$), and hence $g_t$ (Eq.~\ref{eq:gt}).
Then together with $\chi_0$, we obtain the total variational energy (Eq.~\ref{eq:E}) per site explicitly given by
\eqa E = g_t E_K^0 + \frac{U}{2} \sum_k \left(k-\frac{N}{2}+\frac{V}{U}\right)^2 q_k.\eea
where $E_K^0=-N\zeta t \chi_0/2$.
Finally, the energy $E$ needs to be optimized with respect to $(f,g)$ or equivalently $(\Delta_c,g)$.

\subsection{SU(10) case}

\begin{figure}
\centering
\includegraphics[width=1\linewidth]{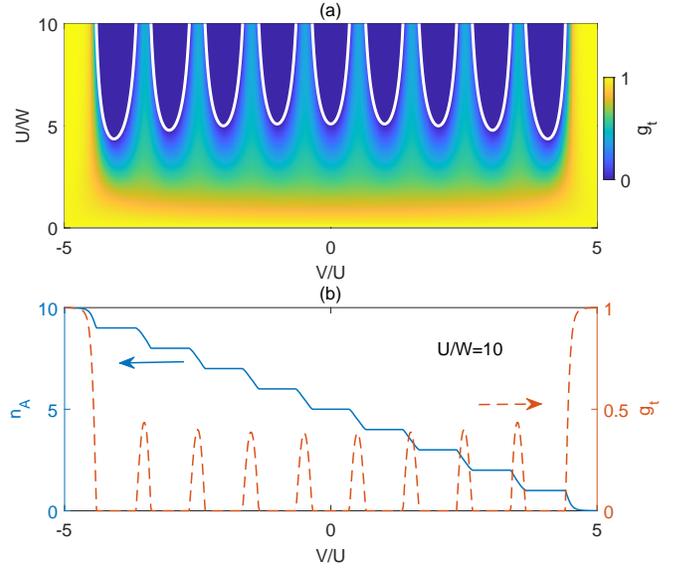}
\caption{Results of the ionic SU(10) Hubbard model. (a) Band renormalization factor $g_t$ versus Hubbard $U$ and ionic potential $V$. The color encodes the value of $g_t$. The white curves enclose Mott lobes with $g_t=0$. (b) Fermion density $n_A$ on the A-sublattice (solid, left scale) and $g_t$ (dashed, right scale) versus $V/U$ for $U/W=10$.}
\label{fig:su10}
\end{figure}

Let us take SU(10) as an example.
In Fig.~\ref{fig:su10}(a), we present the hopping renormalization $g_t$ versus $U$ and $V$.
We find $g_t$ is suppressed by $U$ and drops to zero for large enough $U$ above a critical value $U_c$.
Interestingly, the regimes with $g_t=0$ form different Mott lobes, enclosed by the boundaries shown as white curves (to be calculated analytically in the next section).

For $U/W=10$, we plot the fermion density $n_A$ on the A-sublattice (solid line, left scale) as a function of $V/U$ in Fig.\ref{fig:su10}(b).
For comparison, $g_t$ is also plotted (dashed line, right scale).
As the ionic potential $V$ continuously varies, $n_A$ shows a staircase behavior.
Within each plateau, $n_A=m$ is quantized to the nearest integer of $n_0=\frac{N}{2}-\frac{V}{U}$ and $g_t=0$, where charge fluctuations are completely frozen.
Between neighboring plateaus (Mott lobes), $n_A$ changes continuously between two neighboring integers and meanwhile $g_t$ is nonzero.
{In this region, the system is a band insulator with staggered charge density wave as long as $m\ne N/2$, in which there is no gapless excitation but the charge density $n_A$ as a property of the ground state can be continuously tuned by the staggered ionic potential $V$. In contrast, the uniform part of the charge density (averaged over both sublattices) does not change with $V$, as indicated above.}

\begin{figure}
\centering
\includegraphics[width=0.7\linewidth]{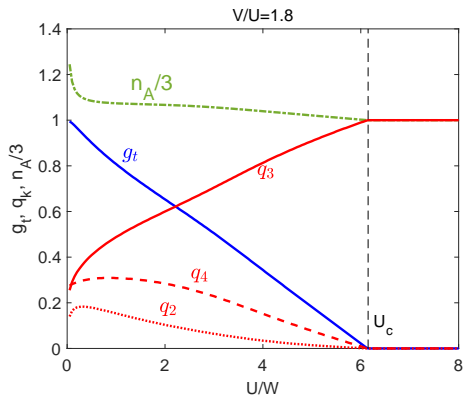}
\caption{Band renormalization factor $g_t$, average of the $k$-tuple projection operator $q_k$ and fermion number $n_A$ versus $U$ for $V/U=1.8$, corresponding to $m=3$ and $\delta=0.2$, in the case of SU($10$). For clarity, only $q_m$ and $q_{m\pm1}$ are plotted.}
\label{fig:su10-Udep}
\end{figure}

In Fig.~\ref{fig:su10-Udep}, we show the $U$-dependence of $g_t$, $n_A$, $q_{m}$ and $q_{m\pm1}$ with a fixed $V/U=1.8$ corresponding to $m=3$.
It is seen that $g_t$ drops continuously with $U$ from $1$ at $U=0$ to $0$ at $U_c$ and maintains at zero for $U>U_c$.
The average fermion number $n_A$ is found to vary continuously with $U<U_c$ but quantized to $m$ when $U\ge U_c$.
The most direct way to see the local moment formation is through $q_m$ which increases with $U$ from the free limit value (given by Eq.~\ref{eq:qk0}) at $U=0$ to $1$ when $U\ge U_c$.
Meanwhile, $q_{k\ne m}$ drops to zero at $U_c$ (for clarity, only $q_{m\pm1}$ are plotted), which of course is a natural consequence of the normalization condition $\sum_k q_k=1$.
Therefore, different $m$-tuple moments are well established inside these Mott insulating phases.

The phase outside of the Mott lobes are characterized by nonzero $g_t$, which in fact is a band insulator (except $V=0$) caused by the ionic potential in our bipartite lattice, although there is a renormalizaiton of the quasiparticle excitations.
Therefore the phase transitions here from $g_t\ne0$ to $g_t=0$ are not the usual metal-insulator transitions but from the band insulator to the Mott insulator.
It is an interesting question to ask whether the band gap closes to generate a metallic phase at or near the phase transition \cite{Garg_PRL_2006, Kancharla_PRL_2007, Garg2019}.
Our answer to this question is actually bilateral:
the effective excitation gap for the quasiparticles (under projection) is given by $g_t\Del_c$, which vanishes as the Mott limit is approached, but at the same time the quasiparticle weight also vanishes.

\subsection{Mott transitions}

\begin{figure}
\centering
\includegraphics[width=1\linewidth]{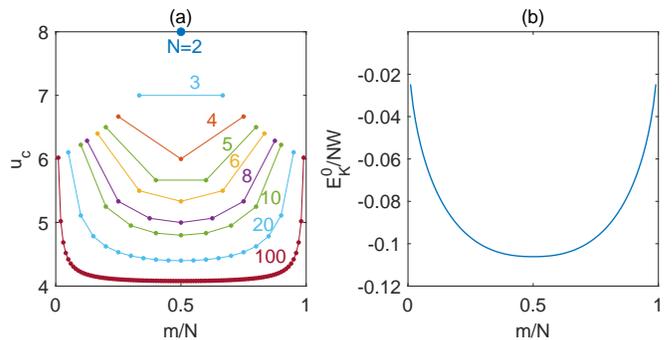}
\caption{(a) The universal function $u_c=U_c/|E_k^0|$ for $\delta=0$ versus $m/N$ for a series of $N$ up to $100$. (b) The kinetic energy per site per flavor $E_K/N$ versus $m/N$.}
\label{fig:ucK0}
\end{figure}

We now try to obtain the critical $U_c$ analytically for the Mott transitions.
Near the Mott lobe labeled by $m$, we have found $q_m$ approaches $1$ and all other $q_{k\ne m}$ are small and linearly drop to zero at $U_c$ as seen from Fig.~\ref{fig:su10-Udep}.
Therefore, it is reasonable to assume
\begin{align}
q_k=(1-\epsilon_--\epsilon_+)\delta_{km} + \epsilon_-\delta_{k,m-1}+\epsilon_+\delta_{k,m+1} ,
\end{align}
which satisfies the normalization condition $\sum_k q_k=1$ and gives the fermion density $Nf=\sum_k kq_k=m+(\eps_+-\eps_-)$.
The hopping renormalization Eq.~\ref{eq:gt} in this approximation is given by
\begin{align}
g_t = \frac{1-\epsilon_--\epsilon_+}{q_{m0}}\left( q_{m-1,0}^{\hat{a}}\sqrt{\frac{\epsilon_-}{q_{m-1,0}}} + q_{m0}^{\hat{a}}\sqrt{\frac{\epsilon_+}{q_{m+1,0}}} \right)^2 ,
\end{align}
such that the total energy per site in the projected state becomes
\begin{align}
E=-g_t|E_K^0|+\frac{U}{2}(\epsilon_-+\epsilon_+)+\frac{U}{4}\delta (\epsilon_+-\epsilon_-) ,
\end{align}
where we defined
\begin{align}\delta=n_0-m=\frac{N}{2}-\frac{V}{U}-m\end{align} as the charge frustration, or the deviation of $N/2-V/U$ away from an integer $m$.
Requiring $\partial E/\partial\epsilon_\pm=0$ in the limit of $\epsilon_{\pm}\to0$, we obtain $U_c=u_c|E_K^0|$, with a universal function $u_c$ independent of the details in the kinetic part of the Hamiltonian,
\begin{align}
u_c= \frac{2}{1-2\delta}\frac{q_{m0}^{\hat{a}2}}{q_{m0}q_{m+1,0}} + \frac{2}{1+2\delta} \frac{q_{m-1,0}^{\hat{a}2}}{q_{m-1,0}q_{m0}} . \label{eq:uc}
\end{align}
Using the expressions for $q_{k,0}$ (Eq.~\ref{eq:qk0}) and $q_{k,0}^{\hat{a}}$ (Eq.~\ref{eq:qk0a}), it can be shown that $U_c$ is automatically invariant under the particle-hole transformation $m\leftrightarrow N-m$ and $f\leftrightarrow 1-f$.
[We note that Eq.~\ref{eq:uc} can also be applied to the non-staggered case.
The only exception is the value of $\chi_0$ (and hence $E_K^0$), to be obtained in a uniform potential which in turn describes a metal.]
From Eq.~\ref{eq:uc}, $u_c$ is found to depend strongly on the charge frustration $\delta$.
As $\delta\to\pm 1/2$ (maximally charge frustrated), $u_c\to\infty$, as seen in Fig.~\ref{fig:su10}(a). This means the Mott transition cannot be reached in this case.
For $\delta=0$, instead, we obtain finite $u_c$, which we plot as a function of $m/N$ in Fig.~\ref{fig:ucK0}(a).
In the case of $N=2$, the Brinkman-Rice result $u_c=8$ is recovered \cite{Brinkman_PRB_1970}.
For larger $N$, $u_c$ is reduced but always larger than $4$.
For a given $N$, $u_c$ increases quickly as $m$ approaches $1$ or $N-1$ but is always smaller than $8$.

\begin{figure}
\centering
\includegraphics[width=1\linewidth]{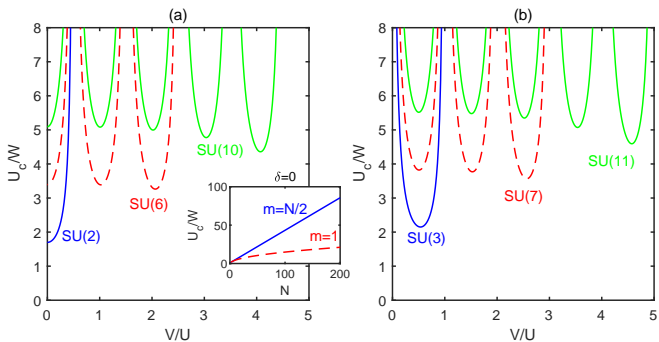}
\caption{(a) The critical value $U_c$ versus $V/U$ for $N=2$ (solid), $6$ (dashed) and $10$ (dash-dotted), respectively. The Mott lobes are labeled by $m$. (b) $U_c/W$ for $\delta=0$ versus $N$ for $m=N/2$ (solid) and $m=1$ (dashed). }
\label{fig:Uc}
\end{figure}

To proceed, we also need $|E_K^0|$ to obtain $U_c$.
For the Bethe lattice, we plot the bare kinetic energy $E_K^0$ per site per flavor with respect to $m/N$ in Fig.~\ref{fig:ucK0}(b).
Clearly, as $m/N$ approaches $0$ or $1$, $|E_K^0|$ drops to zero.
Combining $u_c$ and $|E_K^0|$, we obtain $U_c$.
For the cases of $N=2$, $6$ and $10$, respectively, we plot the results of $U_c$ versus $V/U$ in Fig.~\ref{fig:Uc}(a).
The result of $N=10$ is also plotted in Fig.~\ref{fig:su10}(a) for comparison.
{Similar calculations can be performed on odd $N$ as shown in Fig.~\ref{fig:Uc}(b) for $N=3$, $7$ and $11$, respectively.}
Note that for a fixed $N$ (e.g., $N=10$), $U_c$ drops slightly as $m$ approaches $1$. This is the combined effect of the corresponding increase of $u_c$ [see Fig.~\ref{fig:ucK0}(a)] and decrease of $|E_K^0|$ [see Fig.~\ref{fig:ucK0}(b)].

For the $N$-dependence of $U_c$ with $\delta=0$, we show two representative results of $m=N/2$ and $m=1$ in the inset of Fig.~\ref{fig:Uc}(a).
A perfect linear dependence $U_c$ versus $N$ for large $N$ is seen for $m=N/2$ (solid line) since $u_c$ approaches a constant $4$ from Fig.~\ref{fig:ucK0}(a), and $|E_K^0|\propto N$ from Fig.~\ref{fig:ucK0}(b).
We also find the scaling of $U_c\propto N$ for any fixed $m/N$ (not shown).
But for a fixed $m$, e.g. $m=1$ as shown by the dashed lines in the inset of Fig.~\ref{fig:Uc}(a), $U_c$ does not linearly depend on $N$ any more.
This is because $m/N$ decreases toward zero as $N$ increases to infinity, and thus $|E_K^0|/N$ does not maintain a fixed value but drops to zero.
Therefore, the linear scaling of $U_c=u_c|E_K^0|\propto N$ breaks down to a sublinear behavior.

\section{spin description of the Mott insulating states}

We have found the conditions for different Mott insulating states in which different types of local moments are formed and charge degrees of freedom are frozen.
The low energy effective theory inside these Mott lobes should be described by these local moments, or equivalently the SU($N$) ``spins''.

Given the ground state with $m$ fermions on the A-sublattice and $(N-m)$ fermions on the B-sublattice, we can perform a second order perturbation with respect to the kinetic Hamiltonian $H_t$, to obtain an effective Hamiltonian in the low energy sector,
\begin{align}
H=\frac{4t^2}{(1+\delta^2)U}\sum_{\<ij\>}\sum_{ab}c_a(i)^\dag c_b(i) c_b^\dag(j) c_a(j) ,
\end{align}
subject to $n_i=\sum_a c_a(i)^\dag c_a(i)= m$ on the A-sublattice and $n_i=N-m$ on the B-sublattice.
This restriction suggests to define spin operators $S_{ab}(i)$ on site-$i$ as
\begin{align}
S_{ab}(i)=c_a^\dag(i) c_b(i)-\frac{n_i}{N}\delta_{ab} ,
\end{align}
such that the traceless condition ${\rm Tr}~S(i)=0$ is satisfied.
Further, it can be checked that these $S_{ab}$ satisfy the SU($N$) algebra:
\begin{align}
[S_{ab},S_{cd}]=\delta_{bc}S_{ad}-\delta_{ad}S_{cb}.
\end{align}
Using these spin operators, the above Hamiltonian can be rewritten as the SU($N$) Heisenberg model
\begin{align}
H=J\sum_{\<ij\> ab}S_{ab}(i)S_{ba}(j),
\end{align}
where $J=\frac{4t^2}{(1+\delta^2)U}$. Since $U\sim U_c$ here should be proportional to $N$, the above Hamiltonian has a natural large-$N$ limit.
The SU($N$) Heisenberg model has been widely studied in the literature, as a mathematical generalization of the SU(2) Heisenberg model \cite{Affleck_PRL_1985}.
In this work, we have shown its relation to the ionic SU($N$) Hubbard model.

The above SU($N$) Hubbard or Heisenberg model supports an antiferromagnetic ground state with the Neel order.
To represent the Neel order, we may select a specific spin axis, e.g., one of the diagonal Cartan base,
\begin{align}
L_m(i)=\sum_{a} \ell_{m}^a c_a^\dag(i) c_a(i),
\end{align}
with $\ell_m^a=\frac{1}{m}$ for $a\le m$ and $-\frac{1}{N-m}$ for $a>m$.
The Neel order is then described by $\<L_m(i)\>\sim(-1)^i\cN$, with $m$ flavors of fermions on the A-sublattice and the {\it remaining} $(N-m)$ flavors on the B-sublattice.

Note that even if the local moments $L_m$ are ordered, the state still enjoys an internal symmetry group, $\mathrm{SU}(m)\times \mathrm{SU}(N-m)$, which becomes of merely gauge degrees of freedom if the charge is fully quantized.
The Goldstone modes above the Neel ordered state fall into the Grassmannian manifold $\mathrm{Gr}(N,m) = \mathrm{U}(N)/[\mathrm{U}(m)\times \mathrm{U}(N-m)]$ \cite{MacFarlane_PLB_1979}.
Such fluctuations exchange the flavor content of the local moments without affecting the charge, in analogy to the spin rotation in the SU(2) system.

\section{Conclusion}

In summary, we have developed a Gutzwiller approximation and RMFT for the SU($N$)-symmetric fermionic systems. Applying to the ionic Hubbard model, we find the conjugate local moments, with $m$ fermions on the A-sublattice and $(N-m)$ fermions on the B-sublattice, are well established when the Hubbard $U$ is above a critical value $U_c$. We obtained an analytical solution to $U_c$ wich depends on the bare kinematics and a universal function of $m$, $N$ and the charge frustration $\delta$. For large $N$, $U_c$ is found to depend on $N$ linearly for fixed $m/N$ but sublinearly with fixed $m$ if $N$ is fixed. The local moment formation is accompanied by a peculiar band-insulator to Mott-insulator transition, where the ionic potential and quasiparticle weight are renormalized to zero simultaneously. Inside the Mott insulating phase, the system is effectively described by the SU($N$) Heisenberg model which is widely studied previously in the literature. Our results shed light on the realization of such models in cold atom systems.


{Finally, several remarks on the  Gutzwiller projection are in order. First, it can be improved by including additional Jastrow factors \cite{Edegger_AP_2007}. Second, in one dimension, the Gutzwiller projection is inaccurate or even fails, while  a long-range Jastrow factor alone (without Gutzwiller projection) turns out to be able to capture the Mott insulating state correctly \cite{Jastrow}. Third, even in infinite dimensions, the metal-Mott insulator transition is better described by the Gutzwiller projection followed by a partial Schrieffer-Wolff unitary transformation \cite{BMfail}. The latter two directions are intriguing and even challenge the notion of the Mott state defined by the absence of double occupancy, in the SU($2$) case. It would be interesting to improve our study of the slightly more complicated ionic SU($N$) Hubbard model along similar lines. However, we believe our results for two and higher dimensional ionic models should provide a qualitatively correct picture regarding the multiple transitions from the band insulator to Mott insulator, as well as the order of magnitude of the critical interactions. }



\begin{acknowledgments}
This work is supported by the National Natural Science Foundation of China (under Grant No. 11874205, No. 12274205 and No. 11574134).
\end{acknowledgments}

\bibliography{suN}
\end{document}